\documentclass[notitlepage,floats,aps,nofootinbib,preprintnumbers,superscriptaddress,twocolumn,prl]{revtex4-1}
\setcitestyle{authoryear,round}
\pdfoutput=1
\usepackage{graphicx}
\usepackage{color, graphics}
\usepackage{amsmath,amssymb,amsfonts}
\usepackage{verbatim}   
\usepackage{enumerate}   
\usepackage{subfigure}  
\usepackage{acronym}
\usepackage{hyperref}
\usepackage{mathrsfs}
\usepackage{multirow}
 \usepackage{cancel}
 \usepackage{url}
  \usepackage{slashed}
 \usepackage{hyperref}
 \usepackage{units}

\usepackage{hyperref}
\hypersetup{colorlinks,citecolor=blue}

\definecolor{myblue}{RGB}{0,0,208}
\definecolor{mygreen}{RGB}{29,145,47}
\definecolor{mypurple}{RGB}{164,64,214}
\definecolor{myorange}{RGB}{199,146,32}

\def\beq{\begin{equation}\begin{aligned}}
\def\eeq{\end{aligned}\end{equation}}

\let\OLDthebibliography\thebibliography
\renewcommand\thebibliography[1]{\OLDthebibliography{#1}
  \setlength{\itemsep}{3pt}}

\begin{document}

\title{Dark Matter Freeze-out During Matter Domination}
\author{Saleh Hamdan}
\affiliation{Department of Physics,  University of Illinois at Chicago, Chicago, IL 60607, USA}
\author{James Unwin}
\affiliation{Department of Physics,  University of Illinois at Chicago, Chicago, IL 60607, USA}
\affiliation{DESY, Notkestrasse 85, 22607, Hamburg, Germany}

\date{\today}

\begin{abstract} 

We highlight the general scenario of dark matter freeze-out whilst the energy density of the universe is dominated by a decoupled non-relativistic species. Decoupling during matter domination changes the freeze-out dynamics, since the Hubble rate is parametrically different for matter and radiation domination. Furthermore, for successful Big Bang Nucleosynthesis the state dominating the early universe energy density must decay, this dilutes (or repopulates) the dark matter. As a result, the masses and couplings required to reproduce the observed dark matter relic density can differ significantly from radiation dominated freeze-out.

\end{abstract}

\maketitle


	It is accepted that during Big Bang Nucleosynthesis (BBN) the universe was dominated by radiation, and thus a common assumption is that the universe was radiation dominated from the point of inflationary reheating until the period of BBN. Because of this, models of dark matter (DM) freeze-out typically assume that DM decoupling occurs during radiation domination. However, nothing forbids an early period of matter domination provided that radiation domination is restored prior to BBN. Indeed, any long-lived particle species evolving as matter in the early universe will eventually dominate the energy of the universe, and moduli from string theory and extra-dimensional models \citep{Banks:1993en,de Carlos:1993jw} are good candidates for matter-like (non-relativistic decoupled) states that commonly lead to matter domination in the early universe. 
		
		We consider here a scenario in which the early universe is formed of a thermal bath, comprised of the visible sector states and DM, $X$, along with a decoupled state $\phi$. Immediately after inflationary reheating all of these states are assumed to be  radiation-like, then at some critical temperature $T_\star$ the energy density of $\phi$ begins to evolve as matter. If $\phi$ is sufficiently long-lived, this leads to an early period of matter domination. Notably, if the DM decouples during matter domination this impacts DM freeze-out, since the expansion rate $H$ is different for  matter domination ($H\propto T^{\nicefrac{3}{2}}$) and radiation domination ($H\propto T^{2}$).

			In this work we explore the possibility that DM decoupled whilst the expansion rate scaled as $H\propto T^{\nicefrac{3}{2}}$ when the universe was dominated by a matter-like field $\phi$. To recover radiation domination at BBN $\phi$ must decay and if $\phi$ only decays to Standard Model states (as we will assume), this leads to a dilution of the DM freeze-out abundance. We highlight that the  freeze-out temperature and abundance are distinct from the radiation dominated case, and reproducing the observed DM relic density prefers different DM masses and cross sections. 
			
			This scenario is similar in spirit to \cite{Giudice:2000ex,Gelmini:2006pw,Chung:1998rq} which studied DM freeze-out during inflationary reheating, in which case $H\propto T^{4}$.  However, our work differs from these in several ways, most significantly, these other works consider the case in which the radiation bath is initially negligible, whereas here we consider states decaying at times significantly after inflationary reheating when there is a well established thermal bath. 
			
			\vspace*{-3mm}
\section{Cosmological History}

		A sufficiently long-lived $\phi$ leads to an early period of matter domination. The expansion rate $H$ is different for matter and radiation domination, as can be seen from the Friedmann equation $H^2=\frac{8\pi}{3M_{\rm Pl}^2}\left[\rho_\gamma+\rho_{\phi}+\rho_X\right]$, in terms of $\rho_\gamma$, $\rho_\phi$, $\rho_X$, the energy density of the Standard Model, $\phi$ field, and DM, respectively. We show the Standard Model and DM as separate contributions, however for $T\gtrsim m_X$ they are in thermal contact.
We will parameterise the relative fraction of the total energy at $T=T_\star$ in terms of $r\in[0,1]$, such that for $T_\star>T>m_X$ one has 
\beq
\label{eq:FE}
{\small
\frac{H^2}{H_\star^2}=\frac{g_* r}{g_*+g_X}\left(\frac{a_\star}{a}\right)^{4}+(1-r)\left(\frac{a_\star}{a}\right)^{3}+\frac{g_Xr}{g_*+g_X}\left(\frac{a_\star}{a}\right)^{4}}
\eeq
where $H_\star\equiv H(T_\star)$ and $a_\star\equiv a(T_\star)$ in terms of the FRW scale factor, $g_*$ is the number relativistic Standard Model degrees of freedom and $g_X$ is the DM number of internal degrees of freedom.  $\phi$ starts evolving as $a^{-3}$ when the visible bath temperature is  $T=T_\star$ and in the case that $\phi$ is a decoupled non-relativistic state (as we focus on) 
\beq
\notag
r\equiv\frac{\rho_\gamma+\rho_X}{\rho_\gamma+\rho_X+\rho_\phi}\Bigg|_{T=T_\star}= \left[1+\frac{g_\phi(T_\star)}{g_*(T_\star)+g_X} \left(\frac{m_\phi}{T_\star}\right)^4\right]^{-1}_.
\eeq
Note that one could also consider a field undergoing coherent oscillations, which changes this relation.
Since the thermal bath is radiation-like (assuming entropy is conserved) one has  $(a_\star/a)\simeq(g_*(T)/g_*(T_\star))^{1/4} (T/T_\star)$ and
\beq
H\simeq H_\star\left(\frac{g_*(T)}{g_*(T_\star)}\right)^{\nicefrac{3}{8}}\left(\frac{T}{T_\star}\right)^{\nicefrac{3}{2}}\left[(1-r)+r\left(\frac{T}{T_\star}\right)\right]^{\nicefrac{1}{2}}.
\label{3}
\eeq
For radiation domination ($r\simeq1$) then $H\propto T^{2}$, whereas for $r\ll 1$ the universe is immediately matter dominated and $H\propto T^{\nicefrac{3}{2}}$. Moreover, if the universe remains radiation dominated at $T\simeq T_\star$, it will start evolving towards matter domination after this point. We define matter domination to be when $\phi$ accounts for $50\%$ of the energy of the universe, at which point the temperature of the thermal bath is (this is qualified in the Appendix)
\beq
T_{\rm MD}\equiv T_\star\left(\frac{a_\star}{a(T_{\rm MD})}\right)=\frac{(1-r)}{r}\left(\frac{g_*(T_\star)}{g_*(T_{\rm MD})}\right)^{\nicefrac{1}{4}}T_\star.
\label{22}
\eeq

	The inclusion of the matter-like state $\phi$ allows for a richer range of possibilities, and which scenario is realised depends on the ordering of $T_{\rm FO}$, $T_{\rm MD}$,  and the bath temperature at the point of $\phi$ decays $T_{\Gamma}\equiv T(t\sim\Gamma_\phi^{-1})$:
	
\begin{enumerate}[i).]
	
\item {\em Radiation Domination}:  
If $\phi$ decays prior to DM freeze-out $T_{\Gamma}\gg T_{\rm FO}$, then DM decouples during radiation domination with $H\propto T^2$ and there is no period of $\phi$ matter domination. 

\item {\em Radiation Domination with dilution}:
 DM decouples during radiation domination, again $H\propto T^2$,  but $\phi$ later evolves to dominate the energy density. Decays of $\phi$ dilute or repopulate the DM freeze-out abundance  \citep{Bramante:2017obj}. In this case DM decouples prior to both matter domination and $\phi$ decays: $T_{\rm FO} \gg T_{\Gamma}, T_{\rm MD}$. 
		
\item {\em Matter Domination}:
DM decouples whilst $\phi$ is matter-like and dominates the energy density of the universe, at which point $H\propto T^{\nicefrac{3}{2}}$. This scenario  is realised for $T_{\rm MD} \gg T_{\rm FO} \gg  T_{\Gamma}$.
	
	\end{enumerate}

It is also important to compare these quantities to the reheat temperature after $\phi$ decays\footnote{We use $T_{\rm RH}$ throughout to refer to the heating associated to $\phi$ decays  (rather than inflationary reheating).} $T_{\rm RH}\simeq\sqrt{M_{\rm Pl}\Gamma_\phi}$. If one treats $\phi$ decays as instantaneous, it appears that there is a discontinuous jump in the bath temperature from $T_\Gamma$ to $T_{\rm RH}$. This is an artifact of the sudden decay approximation. Using instead an exponential decay law this jump is absent, rather there is a smooth interpolation and the temperature never rises at any stage  \citep{Scherrer:1984fd}. The effect of $\phi$ decays is seen as a reduction in the rate of cooling. For processes around $T\sim T_{\rm RH}$ the impact of non-instantaneous $\phi$ decays must be taken into account.  Specifically, for DM decoupling with $T_{\rm FO}\sim T_{\rm RH}$, the Boltzmann equations must be modified to include contributions to the number densities due to $\phi$ decays, similar to \cite{Giudice:2000ex}. However, for processes active at temperatures away from this period, instantaneous decay is a fine approximation, and hence we define matter dominated DM freeze-out to be the scenario with $T_{\rm FO}\gg T_{\rm RH}$.

 Whilst scenarios (i) \& (ii) have been discussed in the literature,  to our knowledge, case (iii) remains unstudied. However, matter dominated freeze-out is a very general possibility which readily reproduces the DM relic density and thus we dedicate this letter to the study of case (iii).

\vspace*{-3mm}	
		\section{Matter Dominated Freeze-out}
\vspace*{-3mm}

As in radiation dominated freeze-out, we can define the freeze-out temperature implicitly using the condition $\Gamma_{\rm ann}(T_F)= H(T_F)$ involving the DM annihilation rate $\Gamma_{\rm ann}=n_X^{\rm eq}\langle\sigma v\rangle$, where (in terms of $x\equiv m_X/T$) 
\beq
n_X^{\rm eq}(x)=\frac{g_X}{(2\pi)^{\nicefrac{3}{2}}}m_X^3x^{-\nicefrac{3}{2}}e^{-x}~,
\eeq  
and $\langle\sigma v\rangle$ is the thermally-averaged DM annihilation cross-section. Taking a model independent approach we parameterise the thermally averaged annihilation cross-section as the leading term in an expansion of inverse temperature: $\langle\sigma v\rangle\equiv\sigma_{0}x^{-n}$. We assume throughout that kinetic equilibrium is maintained during freeze-out and for clarity we take $g_*$ to be constant during DM freeze-out, which has no significant impact on the results. 
We next derive $T_{\rm FO}$ using the form of the Hubble rate from eq.~(\ref{eq:FE}), and make the simplifying assumption that the DM contribution can be neglected. This is reasonable since the DM has $g_X\ll g_*$. The point of freeze-out  $x_F\equiv m_X/T_{\rm FO}$ can be found by solving
\begin{equation}
{\small
x_F\simeq\ln{\Bigg[\frac{g_Xm_X^3\sigma_{0}}{(2\pi)^{\nicefrac{3}{2}}H_\star x_\star^{\nicefrac{3}{2}}}\left(\frac{1}{(1-r)x_F^{2n}+ x_\star r x_F^{-1+2n}}\right)^{\nicefrac{1}{2}}\Bigg]}}
\label{xf}
\end{equation}%
with $x_\star\equiv x(T_\star)$. This can be solved numerically for $x_F$, and for $n=0$ ($s$-wave case) an approximate solution is 
	\begin{equation}
	x_F\approx\ln{\Bigg[\frac{g_Xm_X^3\sigma_{0}}{(2\pi)^{\nicefrac{3}{2}}H_\star x_\star^{\nicefrac{3}{2}}}\left(\frac{1}{(1-r)+x_\star r}\right)^{\nicefrac{1}{2}}\Bigg]}.
	\label{xf1}
	\end{equation}
	This approximate solution is in good agreement with exact numerical results.
	 Taking $r\ll1$, and using  $H_\star^2=8\pi^3 g_* T_\star^4/(90M_{\rm Pl}^2)$, gives $x_F$ for matter domination
	\beq
	  x_F^{\rm MD}\equiv x_F\Big|_{r\ll 1}
	\simeq
	\ln{\left[\frac{3}{4\pi^3}\sqrt{\frac{5}{2}}\frac{g_X}{\sqrt{g_*}}\frac{m_X^{\nicefrac{3}{2}}M_{\rm Pl}\sigma_{0}}{\sqrt{T_\star}}\right]}~.
	\label{MD}
	\eeq
	Moreover, the limit $r\rightarrow 1$ reassuringly reproduces the $x_F$ for radiation domination \citep{Scherrer:1985zt}
	\beq
	x_F^{\rm RD} \equiv x_F\Big|_{r\rightarrow 1}
	\simeq
	\ln{\left[\frac{3}{4\pi^3}\sqrt{\frac{5}{2}}\frac{g}{\sqrt{g_*}}m_XM_{\rm Pl}\sigma_{0}\right]}~.
	\label{RD}
	\eeq
	The form of eq.~(\ref{MD}) is distinct from the radiation domination case, in particular, the matter dominated freeze-out temperature exhibits a $T_\star$ dependance. 
	
	For Weak couplings and masses, the characteristic $x_F$ in the radiation domination case is $x_F^{\rm RD}\simeq25$. For DM decoupling during matter dominated, taking $m_X\simeq T_\star\simeq \sigma_{0}^{-\nicefrac{1}{2}}\simeq 10^2$ GeV, freeze-out occurs characteristically at $x_F^{\rm MD}\simeq35$. In both cases these values are only logarithmically sensitive to parameter changes. 
			
	To study the evolution of the DM number density  $n_X$ we use the standard form of the Boltzmann equation
	\begin{equation}
	\dot n_X +3Hn_X=-\langle\sigma v\rangle[n_X^2-(n_X^{\rm eq})^2]~.
	\label{BE}
	\end{equation}
Using $H=-\dot T/T$, the LHS can be expressed in terms of $Y\equiv n_X/s$, where $s=\frac{2\pi^2}{45}g_{* S}T^3$ is the entropy density, as follows: $\dot n +3Hn= -sxH\frac{dY}{dx}$. The expansion rate $H$ here has the form of eq.~(\ref{eq:FE}), and again we simplify by neglecting the DM piece.

Next, we emulate the standard derivation of the freeze-out abundance of non-relativistic DM and write eq.~(\ref{BE}) in terms of $\Delta\equiv Y-Y_{\rm eq}$, as follows
	\begin{equation}
	\Delta'\simeq-Y_{\rm eq}'-\lambda\left(1-r+\frac{rx_\star}{x}\right)^{-\nicefrac{1}{2}}x^{-\nicefrac{5}{2}-n}\Delta[2Y_{\rm eq}+\Delta]~,
	\label{DP2}
	\end{equation}
	with primed variables denoting derivatives with respect to $x$, and
	$\lambda\equiv 2\pi^2g_{* S}m_X^3\sigma_{0}/(45H_\star x_\star^{\nicefrac{3}{2}})$.
	Since we are interested in the freeze-out abundance, we consider the domain $x\gg x_F$. In this regime we have $Y\gg Y_{\rm eq}$, and hence $\Delta\simeq Y$ and also $Y_{\rm eq}'$ may be neglected. Thus the late time  difference $\Delta_\infty\equiv\Delta|_{x\gg x_F}$ follows from
	\begin{equation}
	\Delta'_\infty
	\simeq -\lambda\left(1-r+\frac{rx_\star}{x}\right)^{-\nicefrac{1}{2}} x^{-\nicefrac{5}{2}-n}
	\Delta^2_\infty~.
	\label{DP3}
	\end{equation}
	Solving by separation of variables, we obtain $\Delta_\infty\approx Y_{\rm FO}$ which corresponds to the DM freeze-out abundance
	\begin{equation}
	Y_{\rm FO}\simeq\left(\lambda \int_{x_F}^{\infty}{\rm d}x\left(1-r+\frac{rx_\star}{x}\right)^{-\nicefrac{1}{2}}x^{-\nicefrac{5}{2}-n}
	\right)^{-1}.
	\label{Yinf}
	\end{equation}
	Evaluating the integral on the RHS of eq.~(\ref{Yinf}) with general $n$ and $r$ yields an incomplete $\beta$-function. However, for $n=0$ this can be evaluated to obtain 
	\beq
	Y_{\rm FO}^{n=0}\simeq 
\frac{ x_\star \sqrt{x_F}}{\lambda}\frac{ r}{\sqrt{1-r}}\left(\sqrt{1+y^2}-\frac{1}{y}{\rm sinh}^{-1}(y)\right)^{-1}
	\label{n=0}\eeq
	with $y=\sqrt{r x_\star/(1-r)x_F}$.
Conversely, eq.~(\ref{Yinf}) can be evaluated in the limits of matter and radiation domination to find $Y_{\rm FO}^{\rm MD}$ and $Y_{\rm FO}^{\rm RD}$, respectively. Taking $r\ll1$, for general $n$ one has to leading order
	\beq
	Y^{\rm MD}_{\rm FO}
	=(n+\nicefrac{3}{2}) \frac{x_F^{n+\nicefrac{3}{2}}}{\lambda}
	=3\sqrt{\frac{5}{\pi}}\frac{\sqrt{g_*}}{g_{* S}}\frac{(n+\nicefrac{3}{2})x_F^{n+\nicefrac{3}{2}}}{M_{{\rm Pl}}m_X\sigma_{0}\sqrt{x_\star}}~.
	\label{MDY}
	\eeq
Furthermore, the limit $r\rightarrow1$ reproduces the standard result for $Y^{\rm RD}_{\rm FO}$ \citep{Scherrer:1985zt}. One immediate difference is that $Y^{\rm MD}_{\rm FO}\propto (x_F^{\rm MD})^{n+\nicefrac{3}{2}}$, whereas in the conventional radiation dominated case $Y^{\rm RD}_{\rm FO}\propto (x_F^{\rm RD})^{n+1}$.  Another notable difference is  that  $Y^{\rm MD}_{\rm FO}$ depends on the quantity $x_\star$. Note also that the $n=0$ form of eq.~(\ref{MDY}) agrees with the $r\rightarrow0$ limit of eq.~(\ref{n=0}), as it should.

	    \begin{figure}[t!]
        \centering
        \includegraphics[width=0.38\textwidth]{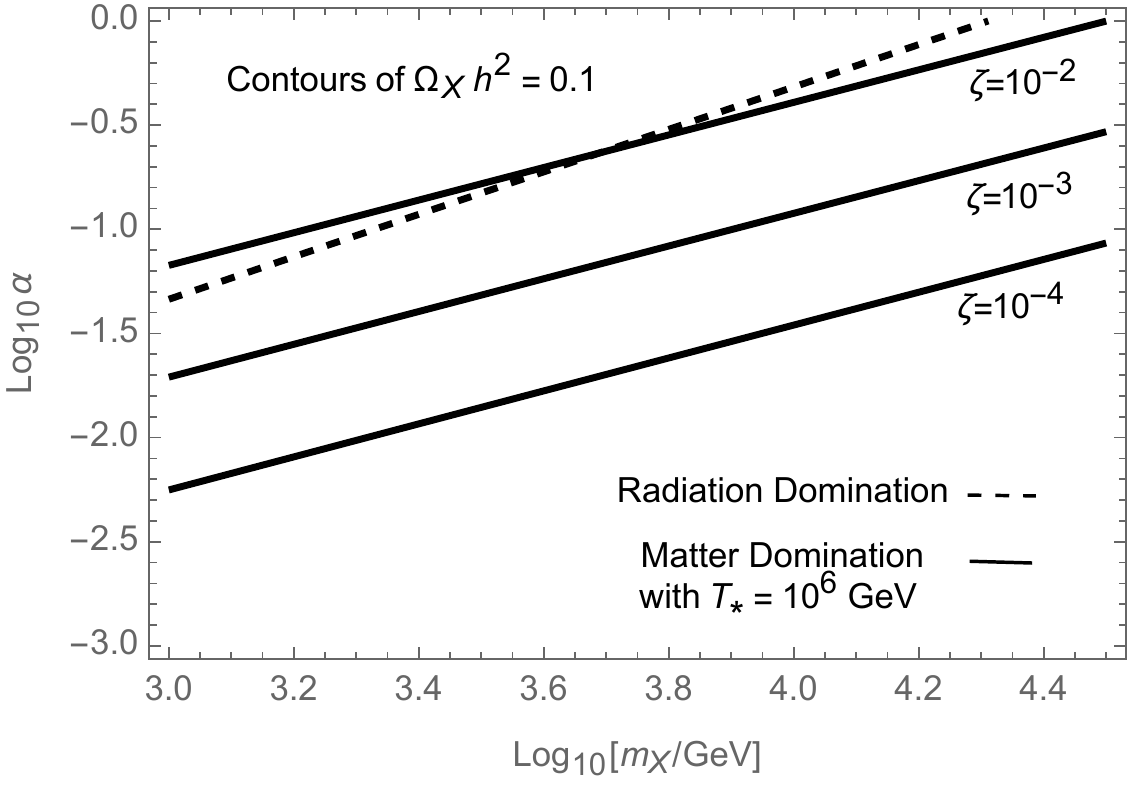}
        \vspace{-3mm}
        \caption{Contours of $\Omega_X h^2=0.1$ for matter (solid) and radiation (dashed) dominated freeze-out scenarios as we vary DM mass and coupling $\alpha$. Matter domination has two extra parameter freedoms, we fix $T_\star=10^6$ GeV and vary $\zeta$.} 
        \label{Fig0}
	\vspace{-3mm}
    \end{figure}

	\section{Dark Matter Relic Abundance}

It is known that the universe was radiation dominated at BBN, hence if DM decouples during $\phi$ matter domination, the $\phi$ states must later decay and restore radiation domination. Specifically, we require that $T_{\rm RH}\simeq\sqrt{\Gamma_\phi M_{\rm Pl}}\gtrsim10$ MeV.
If $\phi$ only decays to Standard Model states (as we assume here) this leads to a dilution of the DM freeze-out abundance. The dilution factor is the ratio of the entropy prior $s_{\rm before}$ and after $s_{\rm after}$ the $\phi$ decays, given by $\zeta\equiv s_{\rm before}/s_{\rm after}$. Thus, it follows that
	\begin{equation}
	Y_{\rm relic}=\frac{n_X}{s_{\rm after}}=\zeta\frac{ n_X}{s_{\rm before}}=\zeta Y_{\rm FO} .
	\end{equation}
	The DM abundance $Y$ can be re-expressed in terms of $\Omega_{\rm DM}=\zeta\times\frac{s_0m_XY_{\rm FO}}{\rho_c}$ by scaling with the critical density $\rho_c$
	and the entropy density today $s_0$.
	Thus for matter dominated freeze-out for $n=0$ ($s$-wave), and writing the annihilation cross section as $\sigma_{0}\equiv\alpha^2/m_X^2$, parametrically
\begin{equation}
{\small
\Omega_{\rm DM}h^2\sim 0.1\left(\frac{m_X}{10^6\rm{GeV}}\right)^{\frac{3}{2}}\left(\frac{0.3}{\alpha}\right)^{2} \left(\frac{T_\star}{10^8\rm{GeV}}\right)^{\frac{1}{2}}\left(\frac{\zeta}{10^{-6}}\right)}
\label{OMDZ}
    \end{equation}
    where we take $g_*, g_{* S}\sim100$, fix $x_F$ to a characteristic value of $x_F\sim35$, and we have normalised eq.~(\ref{OMDZ}) to the observed value \citep{Ade:2015xua}.  In addition to $\zeta$ the value of $x_\star$ presents another new parameter freedom relative to radiation dominated freeze-out. In Figure \ref{Fig0} we give illustrative examples for comparison between radiation and matter dominated freeze-out scenarios.

	To evaluate  $\Omega_{\rm DM}h^2$ we should find the entropy change (see e.g.~\cite{Randall:2015xza}) 
	\begin{equation}
	    \zeta\equiv \frac{s_{\rm before}}{s_{\rm after}}\simeq\left(\frac{g_{* S}(T_\Gamma)}{g_{* S}(T_{\rm RH})}\right)\left(\frac{T_\Gamma}{T_{\rm RH}}\right)^3~.
	    \label{eq:s}
	\end{equation}
Whilst this shows the effect of changing $g_{* S}$, recall we made the simplifying assumption of constant $g_*$  during freeze-out.
The form of $T_\Gamma$ is derived by evolving $T_\star$ using $(a_\star/a_\Gamma)$, where $a_\Gamma$ is defined implicitly by $H(a_\Gamma)\simeq\Gamma_\phi$. 

Assuming that $\phi$ decays after it dominates the energy density, from eq.~(\ref{3}), the ratio of scale factors  can be expressed as an expansion in $H(a_\Gamma)/H_\star$ as follows
{\small \beq
\notag
\frac{a_\star}{a_\Gamma}\approx\frac{1}{(1-r)^{\frac{1}{3}}}\left(\frac{H(a_\Gamma)}{H_\star}\right)^{\frac{2}{3}}-\frac{1}{3}\frac{r}{(1-r)^{\frac{5}{3}}}\left(\frac{H(a_\Gamma)}{H_\star}\right)^{\frac{4}{3}}+\cdots
\eeq}
Thus, for $H(a_\Gamma)\ll H_\star$, to leading order 
 \beq
 T_{\Gamma}=T_\star\left(\frac{a_\star}{a_{\Gamma}}\right)\simeq\left(
  \frac{g_*(T_{\rm RH})}{g_*(T_\star)}\frac{ T_{\rm RH}^{4}}{(1-r)T_\star}\right)^{\nicefrac{1}{3}}~.
  \label{23}
 \eeq
 From eq.~(\ref{eq:s}) \& (\ref{23}), it follows that $\zeta$ is parametrically
\beq
    \zeta  \sim 10^{-6}\times \left(\frac{ 0.01}{1-r} \right) \left(\frac{ T_{\rm RH}}{1~{\rm GeV}} \right) \left(\frac{10^8~{\rm GeV}}{T_\star} \right).
\label{zeta}
\eeq
Here $T_{\rm RH}$ is set to give $\zeta$ and $T_\star$ consistent with eq.~(\ref{OMDZ}).


\vspace{-2mm}
\section{Parameter Space}
\vspace{-2mm}

	Although we have seen the parameter $\zeta$, critical for setting the DM relic density, can take a wide range of values, it is not unbounded.
Indeed, there are several consistency conditions and constraints which must be satisfied for matter dominated DM freeze-out to be viable. Note that in all cases of interest we have that $T_\star\gg  T_{\rm RH}, T_{\rm FO}, T_{\Gamma}$.

\vspace{2mm}{\bf A. Decays of $\phi$.} It is required that $\phi$ decays only after it comes to dominate the energy density of the universe.
From eq.~(\ref{22}) \& (\ref{23}) it is straightforward to verify that $T_{\rm MD} \gg T_{\Gamma}$, as required for a period of matter domination, provided that $T_\star\gg  T_{\rm RH}$.
Additionally, to restore radiation domination after $\phi$ decays, it is necessary that the DM energy density remains small relative to that of $\phi$  immediately prior to $\phi$ decays. Since the DM number density is Boltzmann suppressed, $\rho_\phi\gg \rho_{X}$ at $T=T_{\rm FO}$ and this requirement is typically readily satisfied.

\begin{figure}[t!]
        \centering
        \includegraphics[width=0.4\textwidth]{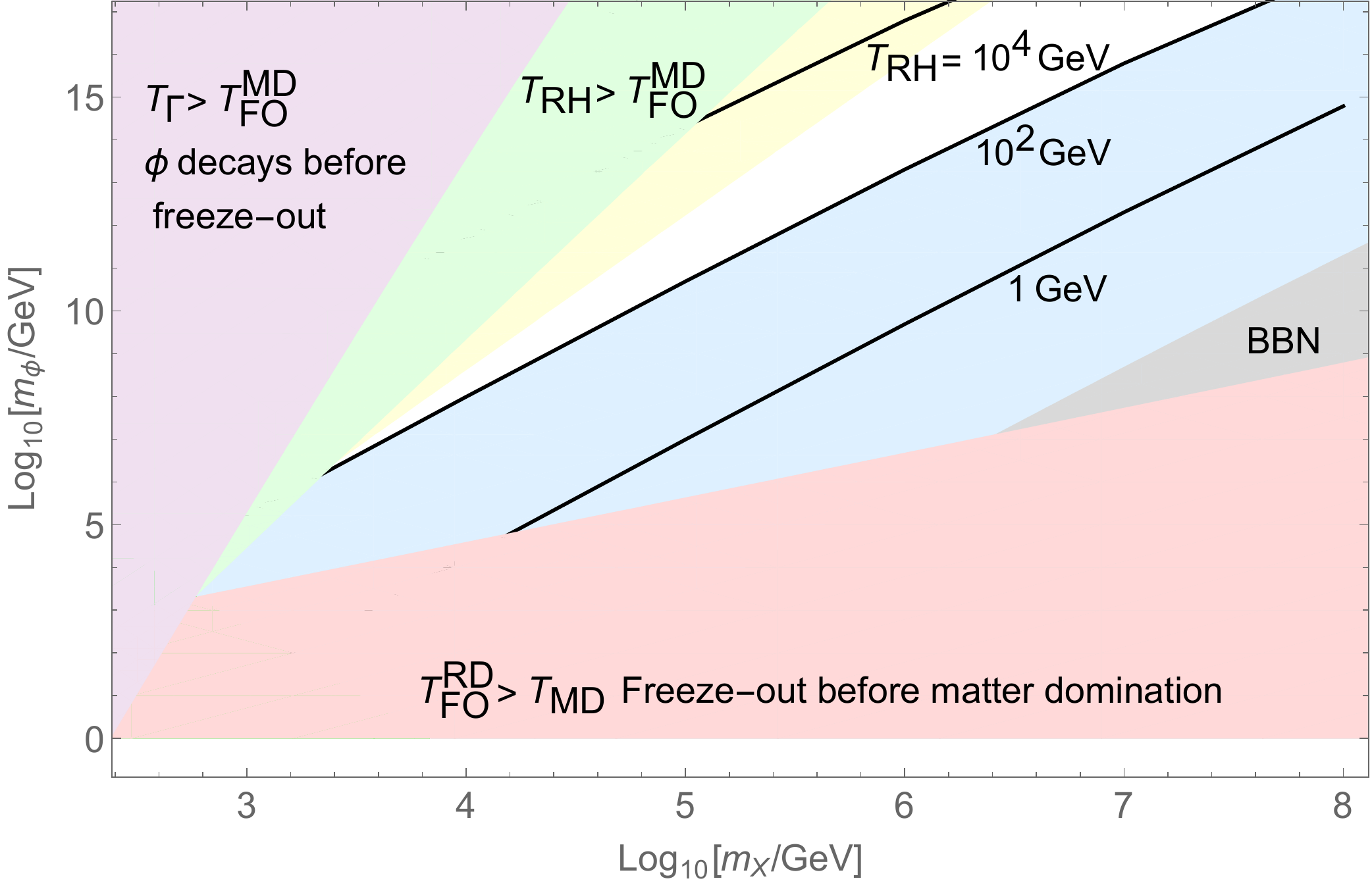}
	\vspace{-4mm}
        \caption{Parameter space of matter dominated DM freeze-out assuming an $s$-wave annihilation cross section $\sigma_{0}\equiv\alpha^2/m_X^2$ for $\alpha=0.1$ with $T_\star\simeq m_\phi$, and thus $r\approx0.99$. We show contours of $T_{\rm RH}$ which give the observed DM relic density. The main constraints are that DM decouples  prior to $\phi$ decays (shaded purple), but after the universe is $\phi$ matter dominated (red). We also require $T_{\rm FO}>T_{\rm RH}$ (green), in the yellow region we expect the instantaneous decay approximation to be less reliable. We shade regions in which the reheat temperature is low: for $T_{\rm RH}\lesssim10$ MeV BBN observables are disrupted (grey) and for $T_{\rm RH}\lesssim100$ GeV baryogenesis is challenging~(blue). }
        \label{Fig1}
	\vspace{-5mm}
    \end{figure}

\vspace{2mm}{\bf B. DM decoupling.}  DM should decouple only after the universe is matter dominated: $T_{\rm MD}\gtrsim T_{\rm FO}^{\rm RD}$ (excluded region shaded red in Fig.~\ref{Fig1}). Thus we should compare the radiation dominated freeze-out temperature, eq.~(\ref{RD}), to  $T_{\rm MD}$ derived in eq.~(\ref{22}).
Further, DM should decouple prior to the period of $\phi$ decays: $T_{\Gamma}\lesssim T_{\rm FO}^{\rm MD}$ (purple in Fig.~\ref{Fig1}) and thus we compare eq.~(\ref{MD}) \& eq.~(\ref{23}). To remain in the matter dominated freeze-out regime we require  $T_{\rm FO}\gg T_{\rm RH}$ (green/yellow in Fig.~\ref{Fig1}). Also, for the DM to be cold it should decouple non-relativistically: $x_F>1$.

\vspace{2mm}{\bf C. Cosmological observations.}
The observed DM relic density must be reproduced ($\Omega_{\rm DM}h^2\sim0.1$), as determined by eq.~(\ref{OMDZ}). Moreover, to avoid conflict with BBN we require $T_{\rm RH}\gtrsim10$ MeV (grey in Fig.~\ref{Fig1}). Both $T_{\rm RH}$ and the DM relic density are controlled by $\zeta$ and $T_\star$, but these requirements can generally be simultaneously satisfied. Higher reheat temperatures are often desirable for baryogenesis mechanisms, e.g.~electroweak baryogenesis typically requires $T_{\rm RH}\gtrsim100$ GeV  (blue in Fig.~\ref{Fig1}).

	{\bf D. Discussion.}
	In Figure \ref{Fig1} we show how the parameter space is constrained by the requirements of A.-C., for  $s$-wave annihilations with $\alpha=0.1$, where we take $T_\star\simeq m_\phi$ (thus $r\approx0.99$), and we neglect changes in $g_*$. In this case the leading constraints are due to the requirement that DM decouples after the universe is $\phi$ matter dominated (shaded red) and $T_{\rm FO}\gg T_{\rm RH}$ (green/yellow). Requiring a reheat temperature above the electroweak  phase transition $T_{\rm RH}\gtrsim100$ GeV (blue) further constrains the parameter space. For the parameter choices made in Figure \ref{Fig1}, the model prefers heavier DM, and can accommodate TeV scale DM.  A more systematic analysis (e.g.~varying $\alpha$ and $r$)  will presented in future work.

	In matter dominated freeze-out, search limits on DM will typically be ameliorated. The bounds are weakened because the DM freeze-out abundance is diluted by $\phi$ decays. Thus DM needs smaller couplings to reproduce the observed DM relic density, relative to the case of radiation domination,  as can be seen in Fig.~\ref{Fig0}. For example, from Fig.~\ref{Fig1}, the observed DM relic density can be obtained for $m_X\sim10^4$ GeV, $\alpha=0.1$,  $T_{\rm RH}\simeq100$ GeV and $m_\phi\simeq 10^{8}$ GeV. In this case the annihilation cross section is $\sigma_0\sim10^{-10} {\rm GeV}^{-2}\sim 10^{-52}~{\rm cm}^2$, and this cross section is far below the characteristic spin independent limits on the scattering cross section of TeV scale DM: $\sigma_{\rm sc}\sim10^{-44}~{\rm cm}^2$ \citep{Aprile:2017iyp}.  We will return to examine carefully constraints from experimental searches in future work within the context of specific models.

\vspace{-3mm}
\section{Summary}
\vspace{-3mm}

In this paper we have explored the prospect that DM decouples whilst the energy density of the universe is dominated by a decoupled non-relativistic species. During matter domination $H\propto T^{\nicefrac{3}{2}}$, this alters the DM freeze-out temperature and abundances compared to radiation dominated freeze-out. To recover successful Big Bang Nucleosynthesis it is required that the state dominating the early universe energy density decays restoring radiation domination, which dilutes (or repopulates) the DM. We have assumed that the $\phi$ decays dominantly to the visible sector such that the DM is not repopulated via $\phi$ decays; direct decays to DM can be forbidden by symmetries or suppressed by small couplings. 

We have derived here the relic abundance of DM following matter dominated freeze-out and entropy dilution, and demonstrated that the correct DM relic density can be realised. Notably, the masses and coupling of DM which reproduce the observed DM relic abundance can be very different from the radiation dominated case.

	The impact of an early period of matter domination on DM has been considered from different perspectives in recent papers, e.g.~\cite{Randall:2015xza}, \cite{Co:2015pka}, \cite{Berlin:2016vnh}, \cite{Bramante:2017obj}, \cite{Mitridate:2017oky}. We also note here an interesting variant, not captured in our list (i)-(iii), in which DM freeze-out occurs whilst the universe is dominated by an energy density redshifting faster than radiation  \citep{DEramo:2017gpl}. An example of this scenario is kinetic energy (`kination') dominated, see e.g.~\citep{Redmond:2017tja}. 
	
	We have outlined the principles of matter dominated DM freeze-out using a model independent approach. In subsequent publications we will consider aspects of model building and examine the phenomenology and search prospects in the context of specific models motivated by common beyond the Standard Model frameworks.

\vspace{2mm}
{\bf Acknowledgements.}
We thank Joe~Bramante, Kai~Schmidt-Hoberg, Jakub Scholtz, and Sebastian~Wild for useful comments and discussions. JU is partially supported by the Alexander von Humboldt Foundation.


\vspace{-5mm}
\subsection*{Appendix: Matter Domination Criteria}
\vspace{-2mm}

	The point of matter domination $T_{\rm MD}$ can be reasonably defined as when matter constitutes roughly $50\%$ of the energy density of the universe. A more precise definition for $T_{\rm MD}$ is when the radiation component of the Friedmann equation is negligible and thus $H\propto T^{\nicefrac{3}{2}}$.
Specifically, from the Friedmann equation for $H/H_\star\ll 1$
\small{\begin{equation}
\notag
\frac{a_\star}{a}\approx\frac{1}{(1-r)^{1/3}}\left(\frac{H}{H_\star}\right)^{2/3}-\frac{1}{3}\frac{r}{(1-r)^{5/3}}\left(\frac{H}{H_\star}\right)^{4/3}+\cdots
\end{equation} }
Matter is the dominant contributor to the Friedmann equation when the second term can be neglected. We parameterise the relative size of the first and second terms by a factor $\beta$, then we define $H_{\rm MD}\equiv H(T_{\rm MD})$ via
\begin{equation}\label{eq:beta}
\frac{\beta}{(1-r)^{1/3}}\left(\frac{H_{\rm MD}}{H_\star}\right)^{2/3}\equiv\frac{1}{3}\frac{r}{(1-r)^{5/3}}\left(\frac{H_{\rm MD}}{H_\star}\right)^{4/3}~.
\end{equation}
For instance, with $\beta=1/3$ the second term is $30\%$ of the first term. This can be solved explicitly for $H_{\rm MD}$ 
\begin{equation}\label{eq:hmd}
\frac{H_{\rm MD}}{H_\star}\equiv(1-r)^{2}\left(\frac{3\beta}{r}\right)^{3/2}~.
\end{equation}
Thus the fraction of energy in $\phi$ at $T=T_{\rm MD}$ is
\beq
\notag
\frac{\rho_\phi}{\rho_\gamma+\rho_{\phi}+\rho_X}\Big|_{T=T_{\rm MD}}=\frac{1}{1+(\frac{a_\star}{a(T_{\rm MD})})\frac{r}{1-r}}=\frac{1}{3\beta+1}~.
\eeq
For $\beta=1/3$, matter domination occurs when the $\phi$ matter accounts for 50\% of the energy of the universe, as used in eq.~(\ref{22}) and also in \cite{Bramante:2017obj}.


\end{document}